\documentstyle[sprocl]{article}

\input{psfig}
\def\mycite#1{$\,$\cite{#1}}

\def\be{\begin{equation}}
\def\ee{\end{equation}}
\def\bea{\begin{eqnarray}}
\def\eea{\end{eqnarray}}

\def\ma{m_{\rm a}}
\def\fa{f_{\rm a}}

\def\dalemb#1#2{{\vbox{\hrule height .#2pt
\hbox{\vrule width.#2pt height#1pt \kern#1pt\vrule width.#2pt}
\hrule height.#2pt}}}

\def\tdot{\kern -8.5pt{}^{{}^{\hbox{...}}}}
\def\dotprime{\kern -8.0pt{}^{{}^{\hbox{.}~\prime}}}

\def\lapp{\hbox{$ {     \lower.40ex\hbox{$<$}
                   \atop \raise.20ex\hbox{$\sim$}
                   }     $}  }
\def\gapp{\hbox{$ {     \lower.40ex\hbox{$>$}
                   \atop \raise.20ex\hbox{$\sim$}
                   }     $}  }

\def\marbul{\strut\vadjust{\kern-2pt$\bullet$}}

\mathchardef\less="321C

\def\smallskip{\vskip 10pt}
\def\microeV{\mu\hbox{\rm eV}}
\def\meV{\hbox{\rm meV}}

\begin{document}

\title{COSMIC AXIONS}

\author{E.P.S.~Shellard and R.A.~Battye}

\address{Department of Applied Mathematics and Theoretical Physics \\
University of Cambridge \\ Silver Street, Cambridge, CB3 9EW, U.K.}

\maketitle\abstracts{The current cosmological constraints on 
a {\it dark matter} axion are reviewed.
We describe the basic mechanisms by which axions are created in the early
universe, both in the standard thermal scenario in which axion strings form
and in inflationary models.  In the thermal scenario, the 
dominant process for axion production is through the radiative decay
of an axion string network, which implies a dark matter axion of mass
$\ma \sim 100\,\microeV$ with specified large uncertainties. An inflationary 
phase does not affect this string bound if the reheat temperature is 
high $T_{\rm reh} \gapp \fa$ or, conversely, for $T_{\rm reh}
\lapp \fa$, if the Hubble parameter during inflation is large $H_1>\fa$;
in both cases, strings form and we return to the standard picture with 
a $\ma \sim 100\,\microeV$ dark matter axion.
Inflationary models with $\fa> H_1>T_{\rm reh}$ face strong CMBR constraints
and require `anthropic misalignment' fine-tuning in order to produce 
a dark matter axion; in this case, some inflation models are essentially 
incompatible with 
a detectable axion, while others can be engineered to allow a dark matter
axion anywhere in a huge mass range below $\ma \lapp 1$meV.  We endeavour 
to clarify the sometimes confusing and contradictory 
literature on axion cosmology.
}
  
\section{Introduction}

The axion has consistently remained one of 
the most popular dark matter candidates ever since the early 1980's.   
Unlike many other more exotic particles,
the axion's existence depends only on a minimal extension of the standard
model\mycite{PecQui77,WilWei78} which also solves one of its key 
difficulties---the strong CP problem of QCD.  
In the standard thermal scenario, cosmic axions are
created primarily through the radiative decay of a network of axion 
strings\mycite{Dav86,DavShe89} formed in the early universe at $t\approx10^{-25}$s; 
this network rapidly annihilates when the axion mass `switches on'
at $t\approx 10^{-4}$s leaving no trace except for a background of 
cosmic axions.
If these axions are to be the dark matter of the universe, then their mass 
must be about $\ma \sim 100\mu\hbox{eV}$ (with significant 
uncertainties).\mycite{BatShe94b} On 
the other hand, if an inflationary phase in the early universe eliminates 
all axions strings with a low reheat temperature,
then dark matter axions are created by 
more quiescent mechanisms\mycite{PAD83} and they can have any 
mass below\mycite{SheBat98} $\ma \lapp 1$meV. Astrophysical 
constraints on the axion mass\mycite{Raf97,axreviews} 
suggest that $\ma \lapp 10\,\meV$, so 
a viable parameter window exists for the dark matter axion; the only 
doubt that remains is 
what it actually should weigh.
  
Present large-scale axion search experiments\mycite{Hagetal98,OMY96} are
focussing on a  mass range, $m_a \sim 1\hbox{--}10\,\microeV$,
for a variety of historical and technological reasons.  However, a number 
of proposed experiments may have the sensitivity to look for the 
heavier dark matter
axion of the standard thermal scenario,\mycite{OtherAxExp} which
appears to have a better motivated mass prediction.  This 
axion cosmology overview, then, keeps this missing matter quest to the forefront.
Our primary focus is not on the viable axion mass range, but rather
on cosmological predictions of the mass of a {\it dark matter} axion, that is, 
an axion for which $\Omega_{\rm a}$ is of order unity.  

\section{Axion properties and constraints}

\noindent The standard model of particle physics, based around the Weinberg-Salam 
model for electroweak interactions and QCD for strong interactions, has 
one significant flaw---the strong CP problem: non-perturbative effects 
of instantons add an extra term to the perturbative Lagrangian
and the coefficient of this term, denoted $\theta$, governs the level
 of CP violation in QCD. The absence of any such violation in all observed 
strong interactions imposes a constraint on $\theta$, the most stringent 
being $\theta<10^{-10}$ due to the absence of a neutron electric dipole 
moment.\mycite{neutrondipole} Since the value of $\theta$ is effectively 
arbitrary, one is left with a severe fine tuning problem.

The elegant solution of Peccei and Quinn\mycite{PecQui77} is to allow $\theta$ to become a 
dynamical field which relaxes toward the CP conserving value 
$\theta=0$~\footnote{In fact, $\theta\ne 0$, since there are some CP violating weak 
interactions, and it has been shown that $\theta>10^{-14}$ in the standard
model, although this is highly model dependent,\mycite{MooWil84}.} through the 
spontaneous breaking of a $U(1)$-symmetry. Essentially, $\theta$ becomes
the phase of a complex scalar field $\phi = |\phi|\exp(i\theta)$, which 
takes a non-zero expectation value  at the Peccei-Quinn scale $\fa$,
that is, it has a potential of the general form,
\be
V(\phi) = {\lambda\over 4}(|\phi|^2 - \fa^2)^2 + {f_\pi^2 m_\pi^2}{|\phi|\over \fa}
(1 - \cos \theta)\,.
\label{potential}
\ee
The azimuthal particle excitation in the $\theta$-direction (around the bottom 
of the `Mexican hat') is a pseudo-Goldstone boson\mycite{WilWei78}---the axion field $a = 
\fa\theta$.
Here, the second term in the potential gives a small mass $\ma$ to the axion, 
however,
note that this mass $\ma(T)$ only `switches on' through instanton effects at 
the QCD-scale $\Lambda_{\rm QCD} \approx 200$MeV (that is, $\ma(T)\approx 0$, 
$T>\!>\Lambda_{\rm QCD}$ and $\ma(T)=\ma,\,\hbox{const.}$, 
$T<\!<\Lambda_{\rm QCD}$). This
ensures that the $\phi$-field lies in the true CP-conserving minimum 
$\theta=0$ at late times.
Throughout this discussion, we consider the simplest $N=1$ axion 
which couples to 
a single quark and has a unique vacuum.  Later, we shall comment on the 
$N>1$ axion with
a $\cos N\theta$ potential, noting the implications of the 
additional vacuum degeneracy.

The axion has an extremely small mass $\ma$ which is inversely proportional
to the Peccei-Quinn scale $\fa$,
\be
\ma = 6.2 \times 10^{-6}\hbox{eV} \left ( 10^{12}\hbox{GeV}\over \fa\right)\,.
\label{axionmass}
\ee
The axion couplings to ordinary matter also scale as $\fa^{-1}$, specifically 
to nucleons $g_{\rm aN} = C_{\rm N} m_{\rm N}/\fa$ (${\rm N = n,\,p}$) and 
to photons $g_{\rm a\gamma} = (\alpha/2\pi\fa)(C_\gamma - 1.9)$ where 
$C_{\rm N}$ and $C_\gamma$ are parameters of order unity. These expressions
do not adequately encompass the model-dependencies of axion phenomenology 
which includes, for example, the DFSZ axion with a tree-level coupling to 
the electron and the KSVZ axion with none (refer to a review such as 
ref.~[14]).

Initially, it was supposed 
that the Peccei-Quinn scale was close to the electroweak phase transition
$\fa\sim T_{\rm EW}$, but an exhaustive search of accelerator 
data ruled out this possibility, implying $\fa\gapp 10^{7}{\rm GeV}$. 
However, the 
ensuing disappointment was short-lived because there is no phenomenological 
reason why the Peccei-Quinn scale $\fa$ could not be much higher, even up to 
grand-unification scales. Thus, the `invisible' axion was born,
an extremely light particle with almost
undetectably weak couplings of order $\fa^{-1}$.

Accelerator limits on the new `invisible' axion were soon superseded by 
astrophysical calculations of stellar cooling rates.  Axions, being 
weakly coupled, can escape from the whole volume
of a star and, in certain parameter ranges, this can exceed the usual 
heat loss mechanisms through convection and other surface effects.  
The oft-quoted `red giant limit' on the axion\mycite{redgiant} 
is actually slightly weaker than that obtained from old globular 
cluster stars,\mycite{Raf97} 
$\ma \lapp 0.4
$eV or $\fa \gapp 2\times 10^7$.  
However, more stringent constraints come from supernova 1987a; the 
duration of the observed neutrino signal, associated with nascent 
neutron star cooling, depends on the axion-nucleon 
coupling $g_{\rm aN}$.  Originally, a strong bound $\ma\lapp 1$meV was
proposed,\mycite{axreviews} but this did not adequately reflect the model-dependence of 
axion phenomenology.  More conservative estimates in realistic axion
models yield the overall bound\mycite{Raf97}
\be
\ma ~\lapp~ 10\,\hbox{meV}\,,\qquad  \fa ~\gapp~ 10^9\hbox{GeV}\,.
\label{astrobound}
\ee
Note that significant uncertainties remain for all these astrophysical 
constraints and recent reviews should be 
consulted such as refs.~[7,8]. Interestingly, astrophysics 
seems to have no impact in any cosmological scenario on the viable 
mass range of a dark matter axion. 

\section{Standard thermal axion cosmology}

\noindent The cosmology of the axion is determined by the two 
energy scales $\fa$ and $\Lambda_{\rm QCD}$.  The first important event
is the Peccei-Quinn phase transition which is broken at a high temperature
$T\sim \fa \gapp 10^9$GeV by (\ref{astrobound}).  This creates the axion,
at this stage effectively massless, as well as a network of axion 
strings\mycite{VilEve82} 
which decays gradually into a background cosmic axions.\mycite{Dav86,DavShe89}
One can, in principle, engineer models in which an 
inflationary epoch interferes with the effects of 
the Peccei-Quinn phase transition, but we shall deal with this case in the 
next section, focussing first on this  `standard 
thermal scenario'.  At a much lower temperature $T\sim \Lambda_{\rm QCD}$
after axion and string formation, instanton effects `switch on', the axions
acquire a small mass, domain walls form\mycite{Sik82} between the 
strings\mycite{VilEve82} 
and the complex
hybrid network annihilates in about one Hubble time.\mycite{She86}  
In the cold dark matter scenario of interest here, the remaining 
axions redshift and 
eventually dominate the energy density of the universe 
at equal matter-radiation $t_{\rm eq}$.

There are three possible mechanisms by which axions are produced in the
`standard thermal scenario':  (i) thermal production, (ii) axion string
radiation and (iii) hybrid defect annihilation when $T=\Lambda_{QCD}$.  
Axions consistent with 
the bound (\ref{astrobound}) decouple from 
thermal equilibrium very early at a temperature\mycite{Tur87} $T_{\rm d}\gapp
4\times10^5$GeV;
their subsequent 
history and number density is analogous to the decoupled neutrino,
except that unlike a 100eV massive neutrino, thermal axions cannot hope 
to dominate the universe with $\ma\lapp 10$meV.  Thermal production
can still continue down to about $\Lambda _{\rm QCD}$ but this contribution 
is subdominant in light of the astrophysical bound (\ref{astrobound}).  
We now turn to the two dominant axion production mechanisms, but first we 
address an important historical digression.

\subsection{Misalignment misconceptions}

\noindent The original papers on axions suggested that axion production
primarily occurred, not through the above mechanisms, but 
instead by `misalignment' effects at the QCD phase transition.
\mycite{PAD83,Tur86}
As applied to the thermal scenario, this turned out to be a very considerable
underestimate,
but it is useful to recount the argument and its flaws because they are 
often repeated in the literature.  Before the
axion mass `switches on', the axion field takes random values throughout
space in the range 0 to $2\pi$.  However, afterwards the true minimum 
becomes $\theta=0$, so the field in the `misalignment' picture begins to 
coherently oscillate about this minimum; this homogeneous mode corresponds
to the `creation' of zero momentum axions. Given an initial
rms value $\theta_{\rm i}$ for these oscillations, it is 
relatively straightforward to 
estimate the total energy density in zero momentum axions and compare 
to the present mass density of the universe (assuming
a flat $\Omega=1$ FRW cosmology):\mycite{PAD83,Tur86,Raf98}
\be
\Omega_{{\rm a},{\rm h}}  ~\approx ~
0.2 \,\Delta h^{-2}\;\theta_{\rm i}^2 
{\rm f}(\theta_{\rm i})\,
\bigg{(}{\fa\over 10^{12}{\rm GeV}}\bigg{)}^{1.18} 
\approx ~ 2 \,\Delta h^{-2}\;\theta_{\rm i}^2 
{\rm f}(\theta_{\rm i})\,
\bigg{(}{10^{-6}{\rm eV}\over \ma}\bigg{)}^{1.18}  
\label{homcount}
\ee
where $\Delta\approx 3^{\pm 1}$ accounts for model-dependent axion 
uncertainties,
as well as those due to the nature of the QCD phase 
transition, and $h$ is the rescaled Hubble parameter. 
The function ${\rm f} (\theta)$ is an anharmonic correction\mycite{Tur86} for 
fields near the top of the potential close to  unstable 
equilibrium $\theta\approx\pi$, that is, 
with ${\rm f}(0)=1$ at the base $\theta\approx 0$ and slowly diverging 
approximately as\mycite{SheBat98}
${\rm f}(\theta) \sim - \ln(|\pi-\theta|)$
for $\theta\rightarrow\pi$.
If valid, the estimate (\ref{homcount}) would imply a 
constraint\mycite{PAD83,Tur86} 
$\fa\lapp 10^{12}{\rm GeV}\,,$ $\ma\gapp 5\mu {\rm eV}$ for the anticipated
thermal initial conditions with $\theta_{\rm i} = {\cal O}(1)$ (suggested 
to be $\theta_{\rm i}=\pi /\sqrt{3}$). 

In the thermal scenario, however, the expression (\ref{homcount}) has
three serious shortcomings:  (i)  The axions are not `created' by 
the mass `switch on' at $t=t_{QCD}$, they are already there 
with a specific
momentum spectrum $g(k)$. Dynamical mechanisms prior to this time---thermal 
effects, string radiation and field realignment---will have created this
axion spectrum.  Because correlations cannot be established acausally, there 
will be a lower cutoff to the axion momenta set by the horizon, $k_{\rm min}^{-1}
\sim t$; only inflationary models can have truly zero momentum 
axions.  The actual axion number obtained from $g( k)$ is clearly much 
larger than an rms average in (\ref{homcount}) 
which ignores the true particle content.
(ii) Secondly, this estimate was derived before much stronger topological 
effects were realized, notably the presence of 
axion strings and domain walls. These
nonlinear effects complicate the oscillatory behaviour of $\theta$ considerably,
implying that the homogeneous estimate (\ref{homcount}) is poorly motivated. 
(iii) Finally, from a more general perspective, the expression (\ref{homcount})
with $\fa>\!> 10^{12}$GeV seems to imply that axion mass `switch on' changes the
topology of the universe, taking us to an $\Omega>1$ closed model!  Of course, 
this is not what happens, since the extrapolation breaks down and 
in reality the axions merely dominate the 
universe much earlier than $t_{\rm eq}$; we end up with 
an $\Omega_{\rm a}\rightarrow 1$ model
with an unacceptably low baryon-to-axion 
ratio.\mycite{Lin88}

\subsection{Axion string network decay}

\noindent Axions and axion strings are inextricably intertwined.  Like
ordinary superconductors or superfluid $^4$He, axion models contain 
a broken $U(1)$-symmetry and so there exist vortex-line solutions.  Combine
this fact with the Peccei-Quinn phase transition, which means the field 
is uncorrelated beyond the horizon, and a random network of axion
strings must inevitably form.  An axion string corresponds to a non-trivial
winding from $0$ to $2\pi$ of the axion field $\theta$ around the bottom of the 
`Mexican hat' potential (\ref{potential}).  It is a global string with 
long-range fields, so its energy per unit length $\mu$ has a logarithmic 
divergence which is cut-off by the string curvature radius $R\lapp t$,
that is, 
\be 
\mu \approx 2\pi \fa^2 \ln (t/\delta)\,,
\label{stringenergy}
\ee
where the string core width is $\delta \approx \fa^{-1}$ 
(taking $\lambda \approx {\cal O} (1)$ in (\ref{potential})).  By observing 
the logarithmic dependence in (\ref{stringenergy}),
the axion string might be interpreted as a delocalised `cloud of energy',
but on cosmological scales it is anything but non-local; if we have a string
stretching across the horizon at the QCD 
temperature, then $\ln(t/\delta)\sim 65$ and over 95\% of its 
energy lies within a tight cylinder enclosing only 0.1\% of 
the horizon volume.  To first order, then, the string behaves like 
a local cosmic string, a fact that can be established by a precise
analytic derivation and careful comparison with numerical 
simulations.~\footnote{The fact
that a global string behaves like a local cosmic string, 
with some additional radiative damping, can be mathematically 
proven by renormalizing the string energy and showing that it obeys exactly 
the same Nambu equations of 
motion.\mycite{Wit85,VilVas87,DavShe89,DabQua90,BatShe94c} Numerical simulations
can then be used to establish the quantitative veracity of this analytic
description.\mycite{DavShe89,BatShe94a,BatShe94c}  
However, great care must be taken in extrapolating string field theory simulations
to cosmological scales, because the numerical results inevitably have a 
limited dynamic range
which is over $10^{26}$ times smaller! For example, on these small numerical
length scales global strings are in a regime where radiative damping is 
more than an order of magnitude stronger.  If such numerical results are simply 
extrapolated to cosmology without a precise quantitative mathematical basis for 
describing the string motion, then it might appear that axion strings radiate
rapidly with a high frequency spectrum.\mycite{HarSik87,HagSik91} The consensus
in the literature, however, weighs heavily against such a 
conclusion.\mycite{VilVas87,DavShe89,DabQua90,BatShe94a,BatShe94b,Sak91,BatShe94c} 
Generically, cosmic axion 
strings radiate like a classical source with a spectrum weighted towards
the lowest available harmonics.}

After formation and a short period of damped evolution, the axion string
network will evolve towards a scale-invariant regime with a fixed number of
strings crossing each horizon volume (for a cosmic string 
review see ref.~[33]). 
This gradual demise of the network
is achieved by the production of small loops which oscillate relativistically 
and radiate primarily into axions.
The overall density of strings splits neatly into
two distinct parts, long strings with length $\ell > t$ and a population of 
small loops $\ell < t$,
\be
\rho=\rho_{\infty}+\rho_{L}\,.
\ee
High resolution numerical simulations\mycite{BenBou90,AllShe90} confirm 
this picture of string 
evolution and suggest that the long 
string density during the radiation era is $\rho_\infty \approx 13\mu/t^2$. 
To date, analytic descriptions of the loop distribution have used the 
well-known string `one scale' model, which predicts a
number density of loops defined as $\mu\ell\,n(\ell,t)\,d\ell=\rho_{L}
(\ell,t)d\ell$ in the interval $\ell$ 
to $\ell+d\ell$ to be given by\mycite{BatShe94b} 
\be
n(\ell,t)={4\alpha^{1/2}(1+\kappa/\alpha)^{3/2}
\over (\ell+\kappa t)^{5/2} t^{3/2}}
\,,\label{numdenloop}
\ee 
where $\alpha$ is the typical loop creation size relative to the horizon 
and $\kappa \approx 65/[2\pi \ln (t/\delta)]$ is the loop radiation 
rate. Once formed at $t=t_0$ with length $\ell_0$, a typical loop 
shrinks linearly as it decays into axions $\ell = \ell_0 - \kappa(t -t_0)$.
The key uncertainty in this treatment is the loop creation size $\alpha$, but 
compelling heuristic arguments place it near the radiative backreaction scale, 
$\alpha \sim \kappa$.

String loops oscillate with a period $T=\ell/2$ and radiate
into harmonics of this frequency (labelled by $n$), just like other 
classical sources.  Unless a loop has a particularly 
degenerate trajectory, it will have a radiation spectrum $P_n \propto n^{-q}$
with a spectral index $q>4/3$, that is, the spectrum is dominated by 
the lowest available modes.  Given the loop density (\ref{numdenloop}), we can 
then calculate the spectral number density 
of axions,
\be
{dn_{\rm a}\over d\omega}(t)\approx {350\alpha^{1/2}(1+\kappa/\alpha)^{3/2}
\fa^2\over 
\kappa^{3/2}\omega^2 t^2}
\left[1-\left(1+{\alpha\over\kappa}\right)^{-3/2}\right]\,,
\ee
which is essentially independent of the exact loop radiation spectrum 
provided  $q>4/3$.   From this expression we can integrate over $\omega$
to find the total axion number at the time $t_{\rm QCD}$, that is, when the axion 
mass `switches on' and the string network annihilates. Subsequently,
the axion number will be conserved, so we can find the number-to-entropy ratio
and project forward to the present day.  Multiplying the present number
density by the axion 
mass $\ma$ will give us the overall string contribution to the density of 
the universe:\mycite{BatShe94b}
\be
\Omega_{\rm a,\ell}\approx 11\Delta h^{-2}
\bigg{[}\bigg{(}1+{\alpha\over\kappa}\bigg{)}^{3/2}-1\bigg{]}
\bigg{(}{\fa\over 10^{12}{\rm
GeV}}\bigg{)}^{1.18}\,.\label{stringbound}
\ee
For the expected ratio $\alpha/\kappa \approx 1$, the axion contribution 
arising from direct radiation from long strings is an order
of magnitude smaller than $\Omega_{\rm a,\ell}$, but it slightly reinforces
(\ref{stringbound}).\mycite{BatShe94b}
We note also that $\Omega_{\rm a,\ell}$ is well over an order of magnitude 
larger than the `misalignment' estimate (\ref{homcount}).    
 The key additional 
uncertainty from the string model is the ratio $\alpha / \kappa$, which  
should be clearly distinguished from 
particle physics and cosmological uncertainties inherent in $\Delta$ and $h$,
and appearing in other estimates of $\Omega_{\rm a}$.  With a Hubble parameter 
near $h=0.5$,
the string estimate (\ref{stringbound}) tends to favour a dark matter axion with a mass 
$\ma \sim 100\mu$eV, as we shall discuss in the conclusion.

\subsection{Axion mass `switch on' and hybrid defect annihilation}

\noindent Near the QCD phase transition the axion acquires a mass and network
evolution alters dramatically because domain walls form.\mycite{VilEve82,Sik82} 
Large field variations around the strings collapse into 
these domain walls, which subsequently begin to dominate over 
the string dynamics.  This occurs when the wall 
surface tension $\sigma$ becomes comparable to the 
string tension due to the typical curvature $\sigma\sim\mu/t$. 
The demise of the hybrid string--wall network proceeds rapidly, as
demonstrated numerically.\mycite{She86,PRS89}  The strings frequently intersect and
intercommute with the walls, effectively `slicing up' the network into small
oscillating walls bounded by string loops.  Multiple self-intersections will reduce
these pieces in size until the strings dominate the dynamics again and decay
continues through axion emission.

An order-of-magnitude estimate of the demise of the string--domain wall
network\mycite{Lyt92} indicates that there is an additional contribution
\be
\Omega_{\rm a,dw}\sim{\cal O}(1)\Delta h^{-2}\bigg{(}{\fa\over 10^{12}{\rm GeV}}\bigg{)}^{1.18}\,.
\ee
This `domain
wall' contribution is ultimately due to loops which are created at 
the time $\sim
t_{\rm QCD}$. Although the resulting loop density will be similar 
to (\ref{numdenloop}),
there is not the same accumulation from early times, so it is likely to be
subdominant\mycite{BatShe94b} relative to (\ref{stringbound}). 
More recent work,\mycite{Nag97}
questions this picture by suggesting that the walls stretching between 
long strings dominate and will produce a contribution anywhere in 
 the wide range $\Omega_{\rm a,dw} 
\sim (1\hbox{--}44) \Omega_{\rm a,\ell}$; however, this assertion 
definitely requires further
quantitative study.
Overall, the domain wall
contribution will serve to further strengthen the string  
bound (\ref{stringbound}) on the axion.

We note briefly that it is possible to weaken any axion mass bound through 
catastrophic entropy production between the QCD-scale and nucleosynthesis,
that is, in the timescale range $10^{-4}s \lapp t_{\rm ent} \lapp 10^{-2}s$.  
The idea has been suggested in a number of contexts,\mycite{Tur86,Lyt93} but 
these usually involve the energy density of 
the universe becoming temporarily dominated by an exotic massive particle 
with a tuned decay timescale.

\section{Inflationary scenarios}

\def\Treh{T_{\rm reh}}

\noindent The relationship between inflation and dark matter axions is
enigmatic.  Its significance depends on the magnitude of the Peccei-Quinn scale 
$\fa$ relative to two key inflationary parameters, (i) the reheat temperature
of the universe $T_{\rm reh}$ at the end of inflation and (ii) the Hubble parameter
$H_1$ as the observed universe first exits the horizon during inflation. 
Inflation is irrelevant to the axion if $\Treh\gapp\fa$ because, in this case, the 
PQ-symmetry is restored and the universe returns to the `standard thermal
scenario' in which axion strings form and the estimate (\ref{stringbound}) pertains.
Even if the reheat temperature is low $\Treh<\fa$, however, inflation will 
again be irrelevant if $H_1>\fa$ because axion strings will form towards
the end of inflation and we will return to a modified 
`standard thermal scenario'---as we
shall discuss. 

Thus, for inflation to impact the viability of a dark matter axion we
must have $\fa\gapp H_1\gapp\Treh $, that is, inflation must occur at a low 
energy scale or we must have an extremely light axion.  Inflation, in this
case, essentially makes no prediction as to the axion mass.  Like much 
in axion cosmology, these facts 
are not widely appreciated, so they deserve some  case-by-case unravelling.

\subsection{$H_1<\fa$: Anthropic misalignment and quantum fluctuations}

\noindent In an inflationary model for which $\fa>H_1>T_{\rm reh}$, the 
$\theta$-parameter or axion angle will be set homogeneously 
over large inflationary 
domains before inflation finishes.\mycite{Pi84}  In this case, the whole observable
universe emerges from a single Hubble volume in which this parameter has some
fixed initial value $\theta_{\rm i}$. Because the
axion remains out of thermal equilibrium for large $\fa$, subsequent evolution 
and reheating does not disturb $\theta_{\rm i}$ until 
the axion mass `switches on' at  $T\sim\Lambda_{\rm QCD}$.  Afterwards,
the field begins
to oscillate coherently, because it is  
misaligned by the angle $\theta_{\rm i}$ from the true minimum $\theta=0$. 
This homogeneous mode corresponds to a background of zero momentum axions 
and it is the one  
circumstance under which the misalignment formula (\ref{homcount})
actually gives an accurate estimate of the relative axion density 
$\Omega_{\rm a}$.

By considering the dependence $\Omega_{\rm a,h} \propto \theta_{\rm i}^2$ 
in (\ref{homcount}), we see that inflation models have an intrinsic 
arbitrariness given by the different random magnitudes of $\theta_{\rm i}$ in
different inflationary domains.\mycite{Pi84}  While a large value of $\fa>\!>10^{12}$GeV 
might have been thought to be observationally excluded, it can actually be 
accommodated in domains where $\theta_{\rm i}<\!<1$.  
This may seem highly 
unlikely but, if we consider an infinite inflationary manifold or a 
multiple universe scenario including `all possible worlds', then 
life as we know it would be excluded from those domains with large
$\theta_{\rm i}={\cal O}(1)$ 
because the baryon-to-axion ratio would be too low.\mycite{Lin88}  
Thus, accepting this anthropic selection effect, we have to 
concede that axions could be the dark matter $\Omega_{\rm a} \approx 1$ 
if we live in a domain with
a `tuned' $\theta$-parameter:~\footnote{This is not quite in the spirit of 
the original motivation for the axion!}
\be
\theta_{\rm i} ~\approx~ \Delta^{-1/2} h\, \left(\fa\over10^{12}{\rm GeV}\right)^{-0.6}
~\approx~
0.3\,\Delta^{-1/2} h\,\left(\ma\over10^{-6}\hbox{eV}\right)^{0.6}\,.
\label{anthropic}
\ee
For $\theta_{\rm i} \approx {\cal O}(1)$, this suggests an axion with
$\ma \sim 5\mu$eV ($h=0.5$), though actually inflation makes no 
definite prediction from 
(\ref{anthropic}) beyond specifying $\ma \gapp 10^{-5}$eV. But even 
this restriction is not valid;\mycite{Raf98} if we observe (\ref{homcount}) 
carefully 
we see that we can also obtain a dark matter axion for higher $\ma$ by 
fine-tuning $\theta_{\rm i}$ near $\pi$. The anharmonic term ${\rm f}(\theta)$
with an apparent logarithmic divergence allows $\Omega_{\rm a}\approx 1$ 
for\mycite{SheBat98} 
$|\pi-\theta_{\rm i}| \sim \exp[-1.5 (\fa/10^{11}\hbox{GeV})^{1.2}]$ with 
$\fa \lapp 10^{11}\hbox{GeV}$, that is, for a much heavier axion.

This simple picture is altered considerably
when we include quantum effects.  Like any minimally coupled massless 
field during inflation, the axion will have a spectrum of 
quantum excitations associated with
the Gibbons-Hawking temperature $T \sim H/2\pi$.  This implies the field 
will acquire fluctuations about its
mean value $\theta_{\rm i}$ of magnitude
\be
\delta\theta = H/2\pi\fa\,,
\label{qfluc}
\ee
giving an effective rms value $\theta_{\rm eff}^2 = (\theta_{\rm i}
+\delta \theta)^2$.  Even if our
universe began in an inflationary domain with $\theta_{\rm i}=0$, there will be
a minimum misalignment angle set by $\delta\theta$; this 
implies that we cannot always fine-tune $\theta_{\rm i}$ in (\ref{anthropic}) 
such that $\Omega_a \lapp 1$.  Taking $\theta_{\rm i} > \delta \theta$, we can 
compare equations (\ref{anthropic}) and  (\ref{qfluc}) to constrain $H_1$
roughly as $H\lapp \fa (\fa/10^{12}\hbox{GeV})^{-0.6}$ 
for $\fa \,\gapp \,10^{12}$GeV ($h=0.5$).
For larger axion masses, we instead require $|\pi-\theta_{\rm i}|>\delta\theta$,
so we can similarly constrain $H\lapp \fa \exp 
[-1.5(\fa/10^{11}\hbox{GeV})^{-1.2}]$ for $\fa\lapp 10^{11}\hbox{GeV}$.  We 
illustrate this restriction schematically in fig.~1, showing that dark matter
axions can be compatible with suitably constructed inflation models for 
any $\ma \lapp 1$meV (see ref.~[22]).

The quantum fluctuations $\delta\theta$, in turn, 
will result in isocurvature fluctuations in the axion density\mycite{isocurv} given by  
(\ref{homcount}), that is, $\delta \rho_{\rm a} 
\propto \theta_{\rm eff}^2
\approx \theta_{\rm i}^2+2\theta_{\rm i}\delta\theta+\delta\theta^2$. Such 
density fluctuations will also create cosmic microwave background (CMB) 
anisotropies which are 
strongly constrained, $\delta T/T \sim \delta \rho/\rho \sim 2\delta\theta
/\theta_{\rm eff}\lapp 10^{-5}$.  Combining the fluctuation (\ref{qfluc})
with the $\theta$-requirement for a dark matter axion (\ref{anthropic}), we
obtain another strong constraint on the Hubble parameter during 
inflation\mycite{Lyt90,TurWil91,Lin91}
\be 
H_1~\lapp ~10^9 \hbox{GeV} \left( \fa\over 10^{13}\hbox{GeV} \right)^{0.4} ~\sim~ 
10^9 \hbox{GeV} \left( \ma\over 10^{-6}\hbox{GeV} \right)^{-0.4}\,.
\label{infbound}
\ee
Here, $H_1$ is the Hubble parameter as fluctuations associated with the
time when the 
microwave anisotropies first leave the horizon some 50--60 e-foldings before
the end of inflation.
It is related to the 
vacuum energy at this time by $V_1^{1/4} \simeq (H_1 \times 10^{18}
\hbox{GeV})^{1/2}$.
We conclude from (\ref{infbound}) that inflation in the $\fa>H_1>T_{\rm reh}$
regime  must have a
small Hubble parameter $H_1$ or the dark matter axion must be extremely 
light,~\footnote{Note that if we were not considering a 
specifically dark matter axion satisfying 
(\ref{anthropic}), 
the bound (\ref{infbound}) is considerably weaker.}  as summarized
in fig.~1.

\begin{figure}
\centerline{\psfig{figure=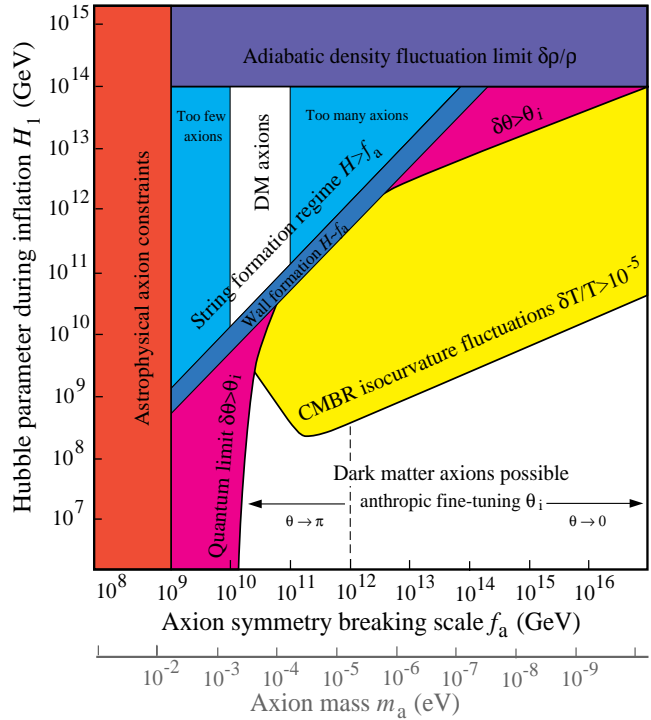,width=3.42in,height=3.8in}}
\caption{{\bf Inflation and the dark matter axion.}  Constraints on the Hubble 
parameter $H_1$ during inflation  and the Peccei-Quinn scale $\fa$ 
and axion 
mass $\ma$.   Only the white areas are compatible with a dark matter
axion $\Omega_{\rm a}\approx 1$  in simple inflation models.  
For $H>\fa$, the string bound 
(\ref{stringbound}) pertains and, 
for $H<\fa$, consistency is due to `anthropic fine-tuning'. 
\label{fig:axion_inflation}}
\end{figure}

Many inflationary models are not very consistent with 
the requirement (\ref{infbound}).
For example, the simplest chaotic inflation models have 
$H_1 \sim 10^{13}\hbox{--}10^{14}$ GeV, implying a truly invisible 
axion with $\ma \sim 10^{-16}$eV and $\fa$ beyond the Planck mass!
However, inflation now has many guises and 
it is possible to engineer 
models with Hubble parameters anywhere in the range 
$H \sim 10^2\hbox{--} 10^{14}$GeV; here, 
the lower bound is set by supersymmetry considerations and the upper bound
by limits on adiabatic density fluctuations (see, for example, 
ref.~[45]).  New inflation 
can have\mycite{ShaVil84}  $H\lapp 10^{9}$GeV with a detectable dark matter axion,
but the low $H$ accentuates its severe initial condition 
difficulties\mycite{Lin91}
(see recent axion inflation reviews such as ref.~[47]).  
Another way to circumvent the CMB constraint (\ref{infbound}) 
is in more complicated inflationary 
scenarios in which the Peccei-Quinn scalar field $\phi$ is
the inflaton\mycite{Pi84,Lin91} or else interacts with it, 
as in a hybrid inflation model.\mycite{Lin91} It is then possible to have 
$|\phi| >\!> \fa$ as the inflaton rolls down to the vacuum, so
the quantum fluctuations in (\ref{qfluc}) will be suppressed~\footnote{Such 
mechanisms might be thought to increase the inflationary upper 
bound $\ma\lapp 1$meV on the dark matter
axion, but the effect is not dramatic because of the exponential 
fine-tuning.\mycite{SheBat98}} 
by the factor $\fa/|\phi|$. 
These inflationary caveats and the lack of any definite axion mass
prediction has led Linde\mycite{Lin91} to conclude that ``the best resolution of 
the uncertainties would be given by an experimental measurement of 
the axion mass. ...  It may also help us to choose between different 
versions of inflation and axion models.''! 

\subsection{$H_1\gapp \fa$: Axion string creation during inflation}

\noindent Even for a low reheat temperature $T_{\rm reh}< \fa$, one can envisage
inflation models with a large Hubble parameter during inflation
$H_1\gapp \fa$.  The quantum fluctuations (\ref{qfluc}) in this case
are sufficient to take the Peccei-Quinn field $\phi$ over the top of the 
potential\mycite{LytSte92} leaving large spatial variations and non-trivial windings 
in $\phi$.  We can interpret this 
as the Gibbons-Hawking temperature `restoring' the PQ-symmetry 
$T_{\rm GH} \gapp \fa$.  
As inflation draws to a close and $H$ falls below $\fa$, these 
fluctuations will become negligible and a string network will form 
along lines where $\phi = 0$.
Provided inflation does not continue beyond this point for more than 
about another 30 e-foldings, we will effectively return to the `standard
thermal' scenario in which axions are produced by a decaying string 
network.  It is effectively as if $T_{\rm reh} > \fa$, so such low reheat 
inflation models are compatible with a dark matter axion\mycite{LytSte92}, $\ma 
\approx 100\mu$eV.  This leaves the additional small upper window for the axion
shown in fig.~1.

More problematic is the case where $H \sim \fa$, or just below, 
for an extended period, say more than $n>30$ e-foldings.  In this case, 
string formation 
may be rare, but the field will still be driven around the Mexican hat potential
by quantum fluctuations ending up with an average total fluctuation 
$\delta \theta
\sim \sqrt n 
H/2\pi \fa\gapp 1$.  This will imprint non-trivial windings 
into the axion field which will appear
as exponentially large domain walls\mycite{LinLyt90} when the axion mass `switches 
on'.  Depending on their size and distribution these are 
potentially incompatible with CMB constraints.\mycite{LinLyt90} A number of 
other complex hybrid scenarios 
can arise when $H_1\sim \fa$, as reviewed in ref.~[48].

Up to this point we have only considered the simplest axion models 
with a unique
vacuum $N=1$, so what happens when $N>1$?  In this case, any strings 
present become attached to $N$ domain walls at the QCD-scale.  Such a 
network `scales' rather than annihilates, and so it is cosmologically
disastrous.\mycite{VilEve82,PRS89}  Consequently, only $\fa>H_1>\Treh$ 
inflationary scenarios are acceptable for $N>1$ and, for these,
constraints like (\ref{homcount}) can be suitably adapted by merely 
rescaling $\fa\rightarrow \fa/N$.

\section{Conclusions}

\noindent We have endeavoured to provide an overview of axion cosmology 
focussing on the mass of a dark matter axion.  First, 
the cosmological axion density has been
calculated in the standard thermal scenario by considering the dominant
contribution from axion strings.  In this case there is, in principle, 
a well-defined calculational
method to precisely predict the mass $\ma$ of a dark matter axion.
For the currently favoured value of the Hubble parameter ($H_0 \approx 
60\rm  \,km\,s^{-1}Mpc^{-1}$), the estimate 
(\ref{stringbound}) predicts a dark matter axion of mass
\be
\ma \approx 200\,\mu\hbox{eV}\,,\qquad \fa \approx 3\times 10^{10}\hbox{GeV}\,,
\ee
where the significant uncertainties from all sources approach an order of 
magnitude.  
The key uncertainty in this string calculation is the parameter ratio
$\alpha/\kappa$, that is, the ratio of the loop size to radiation backreaction 
scale.
Here, like most authors 
we have assumed $\alpha/\kappa\approx1$, however, reducing this 
uncertainty remains a key research goal, though a
technically difficult one.  

Secondly, we have reviewed inflationary axion cosmology showing that 
(i) many inflation models return us to the standard thermal scenario 
with $\ma \sim 100\mu$eV, (ii) some inflation models are essentially 
incompatible with a dark matter axion and (iii),  because of the possibility of 
`anthropic fine-tuning', other inflation models
can be constructed which 
incorporate a dark matter axion mass 
anywhere below $\ma\lapp 1$meV. 

We conclude that, while a dark matter axion might possibly lurk anywhere
in an enormous mass range below $\ma \lapp 1$meV, the best-motivated mass 
for future axion 
searches lies near $\ma \sim 100\mu$eV, a standard thermal scenario 
prediction which is also 
compatible with a broad class of inflationary models.  

\section*{Acknowledgements}

\noindent EPS would like to thank the organisers of COSMO '97 for arranging 
such an interesting meeting.  We have benefitted from 
enlightening discussions with Andrei Linde,
David Lyth and Georg Raffelt.  This work has been supported by PPARC.


\def\jnl#1#2#3#4#5#6{\hang{#1, {\it #4\/} {\bf #5}, #6 (#2).}}


\def\jnlerr#1#2#3#4#5#6#7#8{\hang{#1, {\it #4\/} {\bf #5}, #6 (#2).
{Erratum:} {\it #4\/} {\bf #7}, #8.}}


\def\jnltwo#1#2#3#4#5#6#7#8#9{\hang{#1, {\it #4\/} {\bf #5}, #6 (#2);
{\it #7\/} {\bf #8}, #9.}}

\def\prep#1#2#3#4{\hang{#1 (#2),  #4.}}

\def\myprep#1#2#3#4{\hang{#1 (#2), '#3', #4.}}

\def\proc#1#2#3#4#5#6{\hang{#1 (#2), `#3', in {\it #4\/}, #5, eds.\ (#6).}
}
\def\procu#1#2#3#4#5#6{\hang{#1 (#2), in {\it #4\/}, #5, ed.\ (#6).}
}

\def\book#1#2#3#4{\hang{#1 (#2), {\it #3\/} (#4).}
									}

\def\genref#1#2#3{\hang{#1 (#2), #3}
									}


\def\prl{Phys.\ Rev.\ Lett.}
\def\pr{Phys.\ Rev.}
\def\pl{Phys.\ Lett.}
\def\np{Nucl.\ Phys.}
\def\prp{Phys.\ Rep.}
\def\rmp{Rev.\ Mod.\ Phys.}
\def\cmp{Comm.\ Math.\ Phys.}
\def\mpl{Mod.\ Phys.\ Lett.}
\def\apj{Ap.\ J.}
\def\apjl{Ap.\ J.\ Lett.}
\def\aap{Astron.\ Ap.}
\def\cqg{Class.\ Quant.\ Grav.} 
\def\grg{Gen.\ Rel.\ Grav.}
\def\mn{M.$\,$N.$\,$R.$\,$A.$\,$S.}
\def\ptp{Prog.\ Theor.\ Phys.}
\def\jetp{Sov.\ Phys.\ JETP}
\def\jetpl{JETP Lett.}
\def\jmp{J.\ Math.\ Phys.}
\def\cupress{Cambridge University Press}
\def\pup{Princeton University Press}
\def\wss{World Scientific, Singapore}

\end{document}